\documentclass[aps,graphicx,6pt,showkeys,showpacs,onecolumn]{revtex4}
\usepackage{amsmath}
\usepackage{amscd}
\usepackage{graphicx}
\usepackage{subfigure}
\usepackage{appendix}
\usepackage{amsfonts}
\usepackage{multirow}

\begin{document}

\title[Short Title]{Improving shortcuts to non-Hermitian adiabaticity for fast population transfer in open quantum systems}

\author{Ye-Hong Chen$^{1}$}
\author{Qi-Cheng Wu$^{1}$}
\author{Bi-Hua Huang$^{1}$}
\author{Jie Song$^{2}$}
\author{Yan Xia$^{1,}$\footnote{E-mail: xia-208@163.com}}
\author{Shi-Biao Zheng$^{1}$}

\affiliation{$^{1}$Department of Physics, Fuzhou University, Fuzhou 350002, China\\
             $^{2}$Department of Physics, Harbin Institute of Technology, Harbin 150001, China}


\begin{abstract}
  It is still a challenge to experimentally realize shortcuts to adiabaticity (STA) for a non-Hermitian quantum system
  since a non-Hermitian quantum system's counterdiabatic driving Hamiltonian contains some unrealizable auxiliary control fields.
  In this paper, 
  we relax the strict condition in
  constructing STA and propose a method to redesign a
  realizable supplementary Hamiltonian to
  construct non-Hermitian STA.
  The redesigned supplementary Hamiltonian can be either symmetric or asymmetric.
  For the sake of clearness,
  we apply this method to an Allen-Eberly model as an example to verify the validity of the optimized non-Hermitian STA.
  The numerical simulation
  demonstrates that a ultrafast population inversion could be
  realized in a two-level non-Hermitian system.
\end{abstract}

\pacs {03.67. Pp, 03.67. Mn, 03.67. HK} \keywords{Shortcuts to
adiabaticy; Non-Hermitian Hamiltonian; Two-level system}

\maketitle
\section{Introduction}
Adiabatic methods to manipulate and prepare quantum states are ubiquitous in atomic and molecular physics, nuclear
magnetic resonance, optics, and other fields
\cite{Prl761055,Apb69373,Prl95080502}.
The two main advantages for adiabatic methods are: (i) adiabatic passage is inherently
robust against pulse area and timing errors; (ii) it is
useful in situations where the source and target only interact
via a lossy ``intermediate'' system, as it allows one to use the
mediated coupling without being harmed by the noise.
Despite the advantages, adiabatic methods are necessarily
slow, and hence can suffer from dissipation and noise in
the target and/or source system
\cite{Rmp701003,Rmp7953,Arpc52763,Nc710628,MDSARJpcJpc0308,MVBJpamt09,Prl117080402}.
Therefore, in order to drive a system from a given initial state to a
prescribed final state in a shorter time without losing the
robustness property, many researchers have set their sights on the field of
``Shortcuts to adiabaticity'' (STA) \cite{ETSISMGMMACDGOARXCJGMAmop13,Prl105123003}. 
In the effort to find STA,
several formal and strongly related solutions 
have been proposed, such as, transitionless driving algorithm
\cite{Prl105123003,ETSISMGMMACDGOARXCJGMAmop13,AdCPrl13},
invariant-based inverse engineering \cite{XCJGMPra12,Pra83062116},
using dressed states \cite{Prl116230503}, multiple Schr\"{o}dinger
pictures \cite{Prl109100403}, and so on. Based on these techniques,
in a closed-system scenario, lots of robust protocols for
fast quantum state engineering have been provided in both theory and
practice
\cite{Prl105123003,XCJGMPra12,AdCPrl13,Pra83062116,ETSISMGMMACDGOARXCJGMAmop13,Prl116230503,
Pra13013415,Jpb43085509,Pra85033605,Pra84043434,Pra89012326,CYH,Prl109115703,
Jpb42241001,Njp14013031,Pra84031606Epl9660005,Pra82033430,Prl109100403,Pra89053408,Njp16015025,Pra90060301,Pra08743402,Pra89043408,Np8147,Epl9323001,
Nc712479,Sr515775,Pra93012311}.

On the other hand, it is worth to note that an
increasing interest has been devoted to study non-Hermitian
Hamiltonians in recent years because a non-Hermitian
Hamiltonian, such as, a Hamiltonian obeying $\mathcal{PT}$-symmetric
\cite{Prl805243,Jmp43205}, could produce a faster than Hermitian evolution while keeping the
eigenenergy difference fixed \cite{Prl98040403,Jpa45415304,OE24022847}.
Some non-Hermitian extensions \cite{CJP561007,JPA45444027} also have been done to the Landau-Zener (LZ)
model, which is a standard tool for the description of level-crossing systems.
Recently, STA methods have been generalized to non-Hermitian systems \cite{Pra84023415,Pra8705250289063412,Pra93052109}
and show us a possibility to speed up quantum population transfer without changing coherent control fields.
Unfortunately, the known non-Hermitian STA schemes \cite{Pra84023415,Pra8705250289063412,Pra93052109} require
a well-designed decay rate so that they only remain valid in some particular cases.
For example, in Ref. \cite{Pra84023415} proposed by Ib\'{a}\~{n}ez \emph{et al.},
a limiting condition
$\gamma\ll\Omega_{0}$ is required in order to ensure the supplementary
Hamiltonian is realizable in practice, where $\gamma$ is the decay rate from the
excited state and $\Omega_{0}$ is the amplitude of the Rabi
frequency. That means this scheme \cite{Pra84023415} only remains valid in the
strong-driving regime and the natural lifetime should be large
compared to the duration of the forced decay. Another scheme for
non-Hermitian STA was proposed by Torosov \emph{et al.}
\cite{Pra8705250289063412} through adding a specific imaginary term
in the diagonal elements of the Hermitian original Hamiltonian. As the
additional imaginary term is designed to nullify the
non-adiabatic couplings, the decay rate $\gamma$ should satisfy
a special time-dependent function. 
While in general quantum system such as atomic
system, it is hard or even impossible to design a special
time-dependent $\gamma$ as expected. More recently, Chen \emph{et al.} \cite{Pra93052109}
showed that it was possible to construct STA by nullifying
the transitions from a chosen reference eigenstate to other eigenstaes
while leaving other transitions alone. The idea is promising and has been applied to a
non-Hermitian system. However,
there also exist limiting conditions for the structure and the parameters of the supplementary Hamiltonian.
Therefore, optimizing non-Hermitian STA seems to be imperative in the current situation.

As we know, one of the most common methods in constructing STA in Hermitian systems is the transitionless driving algorithm (also known as counterdiabatic driving) \cite{MVBJpamt09}.
Transitionless driving algorithm shows that one can restrict the system evolution along the instantaneous eigenstates of the original Hamiltonian by adding a supplementary Hamiltonian which can
eliminate the unwanted non-adiabatic transitions \cite{Pra93052109}.
By that analogy, constructing non-Hermitian STA also needs to nullify the non-adiabatic couplings by a supplementary Hamiltonian.
Meanwhile, the supplementary Hamiltonian should be easy to realize
in practice. To find such a supplementary Hamiltonian, we use
the idea of asymmetry transition mentioned in Ref. \cite{Pra93052109} to relax the strict condition for the transitionless
driving algorithm.
Here the asymmetry transition means the transitions between quantum states are asymmetric:
a quantum state $|\psi\rangle$ can be transferred to its orthogonal partner $|\psi_{\perp}\rangle$ with a driving field $\Omega$, while
the transition $|\psi_{\perp}\rangle\rightarrow|\psi\rangle$ can not be realized with the driving field $\Omega^{*}$.
Specifically in this paper, we only prevent the transitions from the reference instantaneous eigenstate $|\phi_{0}(t)\rangle$ to the others $\{|\phi_{n\neq0}(t)\rangle\}$ while
leave the other transitions alone. That is, only the transition $|\phi_{0}(t)\rangle\rightarrow|\phi_{n\neq0}(t)\rangle$ is prevented.
Since the population can not be transferred from $|\phi_{0}(t)\rangle$ to $|\phi_{n\neq0}(t)\rangle$, the system will remain in
$|\phi_{0}(t)\rangle$ all the time if it is initially in $|\phi_{0}(t)\rangle$.
The original Hamiltonian discussed in the present paper is non-Hermitian
so that the supplementary Hamiltonian can be designed either symmetric or asymmetric.
In this way, the problem
caused by the structure of the non-Hermitian part of the supplementary Hamiltonian
could be overcome.
Moreover, by using such a redesigned supplementary Hamiltonian, the non-Hermitian
STA can be constructed even with a relatively large decay rate $\gamma$,
which could be any kind of time-dependent functions. As study
case, we apply the present non-Hermitian STA to perform fast
population inversion in two-level systems.

Noting that the definition of instantaneous eigenstate
populations for a dynamical nonself-adjoint system, i.e.,
non-Hermitian system, is not obvious. The naive direct extension of
the definition used for the self-adjoint case may leads to
inconsistencies; the resulting artifacts can induce a false
population inversion or a false adiabaticity
\cite{Pra89033403,Jpa45415201,Pra94053421}. So, in order to deterministically realize fast
population inversion in non-Hermitian systems, inspired by Refs.
\cite{Pra89033403,Jpa45415201,Pra94053421}, the definition of populations for the instantaneous
eigenstates is modified with a function $f_{n}(t)$ which actually plays
a similar role with the geometric phases. In this case,
the populations of instantaneous eigenstates are (approximatively)
bounded by one, and their sum is (approximatively) one, too.

The rest of this paper is arranged as follows. In Sec. II, we review
the previous non-Hermitian STA based on transitionless driving
algorithm. Then in Sec. III, we show how to use the representation
transformation to construct non-Hermitian STA with a redesigned
supplementary Hamiltonian. As an application example, in Sec. IV, we apply
this method to the popular Allen-Eberly model with a relatively
large decay rate for the realization of fast population inversion.
Conclusion is given in Sec. V.

\section{Transitionless driving algorithm in non-Hermitian systems}
In Ref. \cite{Pra84023415}, Ib\'{a}\~{n}ez \emph{et al.} generalized the transitionless driving algorithm \cite{MVBJpamt09}
for non-Hermitian Hamiltonians to construct non-Hermitian STA.
So, first of all, we would like to give a brief description about Ref. \cite{Pra84023415}.

Non-Hermitian Hamiltonians typically describe subsystems of
a larger system. The basic set of relations and notations for a
non-Hermitian time-dependent Hamiltonian $H_{0}(t)$ with $N$
nondegenerate right eigenstates $\{|n(t)\rangle\}$ and their
biorthogonal partners $\{|\tilde{n}(t)\rangle\}$ ($n=1,2,\cdots,N$),
is given as
\begin{eqnarray}\label{eq1-1}
  H_{0}(t)|n(t)\rangle&=&E_{n}(t)|n(t)\rangle, \cr\cr
  H_{0}^{\dag}(t)|\tilde{n}(t)\rangle&=&E^{*}_{n}(t)|\tilde{n}(t)\rangle.
\end{eqnarray}
$\{|n(t)\rangle\}$ and $\{|\tilde{n}(t)\rangle\}$ satisfy
\begin{eqnarray}\label{eq1-3}
  \langle\tilde{n}(t)|m(t)\rangle&=&\delta_{nm},\cr \cr
  \ \sum_{n}|\tilde{n}(t)\rangle\langle n(t)|&=&\sum_{n}|n(t)\rangle\langle\tilde{n}(t)|=1,
\end{eqnarray}
where $\{\langle\tilde{n}(t)|\}$ and $\{\langle n(t)|\}$ are the left eigenvectors of $H_{0}(t)$ and $H_{0}^{\dag}(t)$, respectively.
Thus, we can write the Hamiltonian and its adjoint as
\begin{eqnarray}\label{eq1-4}
  H_{0}(t)&=&\sum_{n=\pm}|n(t)\rangle E_{n}(t)\langle \tilde{n}(t)|, \cr\cr
  H_{0}^{\dag}(t)&=&\sum_{n=\pm}|\tilde{n}(t)\rangle E_{n}^{*}(t)\langle n(t)|.
\end{eqnarray}
The time-dependent Schr\"{o}dinger equations for a generic state
$|\psi(t)\rangle$ and its biorthogonal partner
$|\tilde{\psi}(t)\rangle$ satisfying
$\langle\tilde{\psi}(t)|\psi(t)\rangle=1$ are
\begin{eqnarray}\label{eq1-4b}
  i\hbar\partial_{t}|\psi(t)\rangle&=&H_{0}(t)|\psi(t)\rangle, \cr\cr
  i\hbar\partial_{t}|\tilde\psi(t)\rangle&=&H_{0}^{\dag}(t)|\tilde\psi(t)\rangle.
\end{eqnarray}

Then, according to transitionless driving algorithm,
the Hamiltonian $H(t)$ that drives the system along the adiabatic paths defined by $H_{0}(t)$
is given as
\begin{eqnarray}\label{eq1-5}
  H(t)=H_{0}(t)+H_{1}(t),
\end{eqnarray}
where $H_{1}(t)$ is the counterdiabatic driving Hamiltonian,
\begin{eqnarray}\label{eq1-6}
  H_{1}(t)&=&i\hbar\sum_{n}[|\partial_{t}n(t)\rangle\langle\tilde n(t)|-\langle\tilde{n}(t)|\partial_{t}n(t)\rangle|n(t)\rangle\langle\tilde{n}(t)|] \cr\cr
          &=&i\hbar\sum_{n}[\sum_{m}|m(t)\rangle\langle\tilde m(t)|\partial_{t}n(t)\rangle\langle\tilde n(t)|
            -\langle\tilde{n}(t)|\partial_{t}n(t)\rangle|n(t)\rangle\langle\tilde{n}(t)|] \cr\cr
          &=&i\hbar\sum_{n\neq m}\langle\tilde m(t)|\partial_{t}n(t)\rangle|m(t)\rangle\langle\tilde{n}(t)|.
\end{eqnarray}
Regarding $\{|n(t)\rangle\}$ and $\{|\tilde{n}(t)\rangle\}$ as the adiabatic basis, $H_{1}(t)$ could be understood
as a matrix in which the $m$th line and $n$th column is $\langle\tilde m(t)|\partial_{t}n(t)\rangle$.
On the other hand, from Eq. (\ref{eq1-3}), we deduce
\begin{eqnarray}\label{eq1-8}
  \langle\tilde{n}(t)|\partial_{t}m(t)\rangle=-\langle\partial_{t}\tilde{n}(t)|m(t)\rangle=-[\langle\tilde{m}(t)|\partial_{t}n(t)\rangle]^{*}.
\end{eqnarray}
Adding this relationship to Eq. (\ref{eq1-6}), we find $H_{1}(t)=[H_{1}(t)]^{\dag}$ holds only when
$\langle\tilde{m}(t)|\partial_{t}n(t)\rangle$ is a real number. While, in a non-Hermitian system,
$\langle\tilde{m}(t)|\partial_{t}n(t)\rangle$ is usually a complex
number. That means, off-diagonal terms in the Hamiltonian in Eq. (\ref{eq1-6}) are not the complex conjugate of each other and
realizing such a Hamiltonian is usually a challenge in practice.



\section{Substitutes of the counterdiabatic driving Hamiltonians}
Known from Ref. \cite{Pra93052109}, for a Hermitian original Hamiltonian $H_{0}(t)$,
the rotation matrixes to transform the quantum system between the interaction frame and the adiabatic frame are given as
\begin{eqnarray}\label{eq1b-1}
  R(t)=\sum_{n}|n(t)\rangle\langle\mu_{n}|\ \text{and}\ R^{\dag}(t)=\sum_{n}|\mu_{n}\rangle\langle n(t)|,
\end{eqnarray}
where $\{|\mu_{n}\rangle\}$ are the bare states.
In the adiabatic frame, the original Hamiltonian $H_{0}(t)$ is written as
\begin{eqnarray}\label{eq1b-2}
  H_{0}^{e}(t)=R^{\dag}H_{0}R-i\hbar R^{\dag}\partial_{t}{R}.
\end{eqnarray}
When $i\hbar R^{\dag}\partial_{t}{R}\ll R^{\dag}H_{0}R$, the reference system behaves
adiabatically, following the eigenstates of $H_{0}(t)$. Also, if we
add a supplementary Hamiltonian $H_{1}^{e}(t)=i\hbar
R^{\dag}\partial_{t}{R}$ into Eq. (\ref{eq1b-2}), the couplings between the
adiabatic basis [the off-diagonal elements in Eq. (\ref{eq1b-2})] will be nullified and the dynamics will be ideally
adiabatic. A general result given in Ref. \cite{Pra93052109} shows that it is in fact not necessary to
nullify all the couplings between the adiabatic basis.
When the transition $|1(t)\rangle\rightarrow|{k}(t)\rangle$ ($k\neq0$) is prevented [the transition $|k(t)\rangle\rightarrow|{1}(t)\rangle$ is allowed],
the system will remain in the reference eigenstate $|1(t)\rangle$ all the time thus STA is constructed.
Further study for Ref. \cite{Pra93052109} shows the
idea fits the requirement of a non-Hermitian system perfectly
because in most of the cases, asymmetry transition only happens in
non-Hermitian systems.

In the non-Hermitian case, the rotation matrixes to transform the
quantum system between the interaction frame and the adiabatic
frame with the relationships
$|\psi^{e}(t)\rangle=\tilde{R}^{\dag}|\psi(t)\rangle$ and
$|\psi(t)\rangle={R}|\psi^{e}(t)\rangle$ read
\begin{eqnarray}\label{eq1b-3}
  R(t)=\sum_{n}|n(t)\rangle\langle\mu_{n}|\ \text{and}\ \tilde{R}^{\dag}(t)=\sum_{n}|\mu_{n}\rangle\langle \tilde{n}(t)|,
\end{eqnarray}
where $\tilde{R}^{\dag}$ and $R$ satisfy $\tilde{R}^{\dag}R=1$. We can accordingly rewrite the original Hamiltonian $H_{0}(t)$ in the adiabatic frame as
\begin{eqnarray}\label{eq1b-4}
  H_{0}^{e}(t)=\tilde{R}^{\dag}H_{0}R-i\hbar \tilde{R}^{\dag}\partial_{t}{R},
\end{eqnarray}
or in the form of a matrix
\begin{eqnarray}\label{eq1b-5}
  H_{0}^{e}(t)=\left(
                    \begin{array}{cccc}
                      E_{1}-i\hbar\langle\tilde{1}(t)|\partial_{t}{1}(t)\rangle & -i\hbar\langle\tilde{1}(t)|\partial_{t}{2}(t)\rangle & -i\hbar\langle\tilde{1}(t)|\partial_{t}{3}(t)\rangle & \cdots \cr\cr
                      -i\hbar\langle\tilde{2}(t)|\partial_{t}{1}(t)\rangle & E_{2}-i\hbar\langle\tilde{2}(t)|\partial_{t}{2}(t)\rangle & -i\hbar\langle\tilde{2}(t)|\partial_{t}{3}(t)\rangle & \cdots \cr\cr
                      -i\hbar\langle\tilde{3}(t)|\partial_{t}{1}(t)\rangle & -i\hbar\langle\tilde{3}(t)|\partial_{t}{2}(t)\rangle & E_{3}-i\hbar\langle\tilde{3}(t)|\partial_{t}{3}(t)\rangle & \cdots \cr\cr
                      \vdots & \vdots & \vdots & \vdots
                    \end{array}
               \right).
\end{eqnarray}
Obviously, if we want to prevent the transition $|1(t)\rangle\rightarrow|k(t)\rangle$,
the supplementary Hamiltonian in the adiabatic frame should be chosen as
\begin{eqnarray}\label{eq1b-6}
  H_{1}^{e}(t)=\left(
                    \begin{array}{cccc}
                      A_{11} & i\hbar\langle\tilde{1}|\partial_{t}{2}\rangle & i\hbar\langle\tilde{1}|\partial_{t}{3}\rangle & \cdots \cr
                      A_{21} & A_{22} & A_{23} & \cdots \cr
                      A_{31} & A_{32} & A_{33} & \cdots \cr
                      \vdots & \vdots & \vdots & \vdots
                    \end{array}
               \right),
\end{eqnarray}
where $\{A_{nm}\}$ are arbitrary coefficients.
The supplementary Hamiltonian in the interaction frame reads $H_{1}(t)=RH_{1}^{e}\tilde{R}^{\dag}$.
Hence, one can choose suitable $\{A_{nm}\}$ to to ensure the Hamiltonian $H_{1}(t)$ is realizable in practice.

\section{Application Example OF Two-level systems}
For the sake of clearness, we take a two-level system as an
example to verify the feasibility of the idea proposed above.
Applying the electric dipole approximation, a laser-adapted
interaction frame, and the rotating wave approximation, the
Hamiltonian, disregarding atomic motion, is given as
\begin{eqnarray}\label{eq2-1}
  H_{0}(t)=\frac{\hbar}{2}
           \left(\begin{array}{cc}
             -\Delta(t) & \Omega_{R}(t) \\
             \Omega_{R}(t) & \Delta(t)-i\gamma(t)
           \end{array}
           \right),
\end{eqnarray}
in the atomic basis $|0\rangle=[1,0]^{t}$, $|1\rangle=[0,1]^{t}$ (the superscript $t$ denotes the transpose).
In Eq. (\ref{eq2-1}), $\Omega_{R}(t)$ is the Rabi frequency, $\Delta(t)$
is the detuning, and $\gamma(t)$ is the decay rate from the excited state. The eigenvalues for this Hamiltonian are
\begin{eqnarray}\label{eq2-2}
  E_{\pm}=\frac{\hbar}{4}[-i\gamma\pm\sqrt{-(\gamma+2i\Delta)^{2}+4\Omega_{R}^{2}}],
\end{eqnarray}
and the eigenstates are
\begin{eqnarray}\label{eq2-3}
  |+(t)\rangle&=&\cos(\frac{\theta}{2})|0\rangle+\sin(\frac{\theta}{2})|1\rangle, \cr\cr
  |-(t)\rangle&=&\sin(\frac{\theta}{2})|0\rangle-\cos(\frac{\theta}{2})|1\rangle,
\end{eqnarray}
where the time-dependent mixing angle $\theta(t)$ is complex and defined by
\begin{eqnarray}\label{eq2-4}
  \tan[\theta(t)]={\frac{\Omega_{R}}{\Delta-i\gamma/2}}.
\end{eqnarray}
The biorthogonal partner Hamiltonian for $H_{0}(t)$ in Eq. (\ref{eq2-1}) reads
\begin{eqnarray}\label{eq2-6}
  H_{0}^{\dag}(t)=\frac{\hbar}{2}
               \left(\begin{array}{cc}
                 -\Delta(t) & \Omega_{R}(t) \\
                 \Omega_{R}(t) & \Delta(t)+i\gamma(t)
               \end{array}
               \right),
\end{eqnarray}
with eigenstates
\begin{eqnarray}\label{eq2-7}
  |\tilde{+}(t)\rangle&=&\cos(\frac{\theta^{*}}{2})|0\rangle+\sin(\frac{\theta^{*}}{2})|1\rangle, \cr\cr
  |\tilde{-}(t)\rangle&=&\sin(\frac{\theta^{*}}{2})|0\rangle-\cos(\frac{\theta^{*}}{2})|1\rangle,
\end{eqnarray}
and the corresponding eigenvalues
\begin{eqnarray}\label{eq2-8}
  E_{\pm}^{*}(t)=\frac{\hbar}{4}[i\gamma\pm\sqrt{-(\gamma-2i\Delta)^{2}+4\Omega_{R}^{2}}].
\end{eqnarray}
With the eigenvectors $\{|n(t)\rangle\}$ and Hamiltonian $H_{0}(t)$,
we may expand the evolution state $|\psi(t)\rangle$ satisfying the Schr\"{o}dinger equation
$i\hbar\partial_{t}|\psi(t)\rangle=H_{0}(t)|\psi(t)\rangle$
as
\begin{eqnarray}\label{eq2-12}
  |\psi(t)\rangle=\sum_{n=\pm}c_{n}(t)|n(t)\rangle.
\end{eqnarray}
It is not hard to find from Eq. (\ref{eq2-12}) that $c_{n}(t)=\langle\tilde{n}(t)|\psi(t)\rangle$, but
$|c_{n}(t)|^{2}$ is not bounded by one, and $\sum_{n}|c_{n}(t)|^{2}\neq 1$.
Thus, it is inappropriate to use $\{c_{n}(t)\}$ to define the probability amplitudes of
eigenstates $\{|n(t)\rangle\}$ \cite{Pra89033403,Jpa45415201,Pra94053421}.
A relatively suitable definition for non-Hermitian eigenstates according to Refs. \cite{Pra89033403,Jpa45415201,Pra94053421}
is given by using states
\begin{eqnarray}\label{eq2-13}
  |\phi_{n}(t)\rangle&=&f_{n}(t)|n(t)\rangle, \cr\cr
  |\tilde{\phi}_{n}(t)\rangle&=&\frac{1}{f_{n}^{*}(t)}|\tilde{n}(t)\rangle,
\end{eqnarray}
which can also constitute a complete, biorthogonal set of eigenstates of $H_{0}(t)$,
where $f_{n}(t)\in\mathbb{C}$ is an arbitrary function.
Adding Eq. (\ref{eq2-13}) into Eq. (\ref{eq2-12}), we have
\begin{eqnarray}\label{eq2-14}
  |\psi(t)\rangle=\sum_{n=\pm}g_{n}(t)|\phi_{n}(t)\rangle,
\end{eqnarray}
where $g_{n}(t)=\frac{c_{n}(t)}{f_{n}(t)}=\langle\tilde{\phi}_{n}(t)|\psi(t)\rangle$. Obviously, when we suitably choose $f_{n}(t)$, it is possible to make $|g_{n}(t)|^{2}$
to satisfy normalization $\sum_{n=\pm}|g_{n}(t)|^{2}=1$. Then, according to Eq. (\ref{eq2-14}), the population for eigenstate $|\phi_{n}(t)\rangle$ can be defined as
$P_{\phi_{n}}=|g_{n}(t)|^2$, where $\{g_{n}(t)\}$ are regarded as the modified probability amplitudes of the
adiabatic basis.


In the following, $\{|\phi_{n}(t)\rangle\}$ ($\{|\tilde{\phi}_{n}(t)\rangle\}$) will be used to replace $\{|n(t)\rangle\}$ ($\{|\tilde{n}(t)\rangle\}$) as the adiabatic basis for further study.
To study the adiabaticity of this two-level non-Hermitian system, we introduce a vector $|\psi^{e}(t)\rangle=[g_{+}(t),g_{-}(t)]^{t}$
to describe the evolution state in the adiabatic frame.
$|\psi^{e}(t)\rangle$ and $|\psi(t)\rangle$ are connected via the rotation matrixes
\begin{eqnarray}\label{eq2-15}
  R(t)=\left(
                 \begin{array}{cc}
                   f_{+}(t)\cos{\frac{\theta}{2}} &\ f_{-}(t)\sin{\frac{\theta}{2}} \cr\cr
                   f_{+}(t)\sin{\frac{\theta}{2}} &\ -f_{-}(t)\cos{\frac{\theta}{2}}
                 \end{array}
            \right), \
            \text{and} \
  \tilde{R}(t)=\left(
                 \begin{array}{cc}
                   \frac{1}{f_{+}^{*}}\cos{\frac{\theta^{*}}{2}} &\ \frac{1}{f_{-}^{*}}\sin{\frac{\theta^{*}}{2}} \cr\cr
                   \frac{1}{f_{+}^{*}}\sin{\frac{\theta^{*}}{2}} &\ -\frac{1}{f_{-}^{*}}\cos{\frac{\theta^{*}}{2}}
                 \end{array}
            \right),
\end{eqnarray}
with relationships $|\psi(t)\rangle=R(t)|\psi^{e}(t)\rangle$ and $|\psi^{e}(t)\rangle=\tilde{R}^{\dag}(t)|\psi(t)\rangle$.
In this case, the Schr\"{o}dinger equation in the adiabatic frame reads
\begin{eqnarray}\label{eq2-16}
  i\hbar\partial_{t}|\psi^{e}(t)\rangle=H_{0}^{e}(t)|\psi^{e}(t)\rangle,
\end{eqnarray}
where
\begin{eqnarray}\label{eq2-17}
  H_{0}^{e}(t)&=&\tilde{R}^{\dag}H_{0}R-i\hbar{\tilde{R}}^{\dag}\partial_{t}{R}  \cr\cr
              &=&\left(
                       \begin{array}{cc}
                         E_{+} & 0  \\
                         0 & E_{-}
                       \end{array}
                 \right)
                 -
                 i\hbar\left(
                       \begin{array}{cc}
                         \frac{\partial_{t}{f}_{+}}{f_{+}} &\ \frac{(\partial_{t}{\theta})f_{-}}{2f_{+}} \\
                         -\frac{(\partial_{t}{\theta})f_{+}}{2f_{-}} &\ \frac{\partial_{t}{f}_{-}}{f_{-}}
                       \end{array}
                 \right).
\end{eqnarray}
The off-diagonal terms in Eq. (\ref{eq2-17}) are the non-adiabatic couplings.
According to Eq. (\ref{eq2-17}), when $|\frac{(\partial_{t}{\theta})f_{-}}{2f_{+}}|\ll |E_{+}-i\hbar\frac{\partial_{t}{f}_{+}}{f_{+}}|$ and
$|\frac{(\partial_{t}{\theta})f_{+}}{2f_{-}}|\ll |E_{-}-i\hbar\frac{\partial_{t}{f}_{-}}{f_{-}}|$,
by solving the Schr\"{o}dinger equation in Eq. (\ref{eq2-16}), we obtain
\begin{eqnarray}\label{eq2-18}
  \left(
       \begin{array}{c}
         g_{+}(t) \cr\cr
         g_{-}(t)
       \end{array}
  \right)
  \simeq
  \left(
       \begin{array}{c}
         g_{+}(t_{0})\cdot\exp[{-i\int_{t_{0}}^{t}\frac{E_{+}(t')}{\hbar}-\frac{i\partial_{t}{f}_{+}(t')}{f_{+}(t')}}dt'] \cr\cr
         g_{-}(t_{0})\cdot\exp[{-i\int_{t_{0}}^{t}\frac{E_{-}(t')}{\hbar}-\frac{i\partial_{t}{f}_{-}(t')}{f_{-}(t')}}dt']
       \end{array}
  \right).
\end{eqnarray}
Obviously, when
${\int_{t_{0}}^{t}[\frac{E_{n}(t')}{\hbar}-\frac{i\partial_{t}{f}_{n}(t')}{f_{n}(t')}]}dt'$ is real,
the result of Eq. (\ref{eq2-18}) shows $|g_{n}(t)|\simeq|g_{n}(0)|$ which means the evolution is adiabatic.
The condition for $f_{n}(t)$ is
\begin{eqnarray}\label{eq2-19}
  \frac{\partial_{t}{f}_{n}(t)}{f_{n}(t)}=\frac{\text{Im}[E_{n}(t)]}{\hbar}+ih_{n}(t),
\end{eqnarray}
where $h_{n}(t)$ is an arbitrary real function.
A simple solution for $f_{n}(t)$ is
\begin{eqnarray}\label{eq2-20}
  f_{n}(t)=\exp\{{\int_{t_{0}}^{t}\frac{\text{Im}[E_{n}(t')]}{\hbar}+ih_{n}(t')dt'}\}.
\end{eqnarray}

To construct shortcuts in the system mentioned above, we need to add supplementary terms $H_{1}^{e}(t)$ in the Hamiltonian described by Eq. (\ref{eq2-17})
to nullify the non-adiabatic couplings.
The simplest choice for $H_{1}^{e}(t)$ is
\begin{eqnarray}\label{eq2-21}
  H_{1}^{e}(t)=i\frac{\hbar}{2}\left(
                          \begin{array}{cc}
                             \varepsilon_{+} & \frac{(\partial_{t}{\theta})f_{-}}{f_{+}} \cr\cr
                            -\frac{(\partial_{t}{\theta})f_{+}}{f_{-}}& \ \varepsilon_{-}
                          \end{array}
                    \right),
\end{eqnarray}
where $\varepsilon_{\pm}$ are undetermined coefficients.
Transforming Eq. (\ref{eq2-21}) back to the interaction frame with $H_{1}(t)=R H_{1}^{e}\tilde{R}^{\dag}$, we obtain
\begin{eqnarray}\label{eq2-22}
  H_{1}(t)=i\frac{\hbar}{2}\left(
                       \begin{array}{cc}
                            \varepsilon_{+}\cos^{2}{\frac{\theta}{2}}+\varepsilon_{-}\sin^{2}\frac{\theta}{2} &\ \
                            \frac{\sin{\theta}}{2}(\varepsilon_{+}- \varepsilon_{-})
                            -\frac{\partial_{t}{\theta}}{2} \cr\cr
                            \frac{\sin{\theta}}{2}(\varepsilon_{+}- \varepsilon_{-})
                            +\frac{\partial_{t}{\theta}}{2} & \ \ \varepsilon_{+}\sin^{2}{\frac{\theta}{2}}+\varepsilon_{-}\cos^{2}\frac{\theta}{2}
                       \end{array}
                    \right).
\end{eqnarray}
As we can find, the practical realization of this Hamiltonian is not straightforward since the off-diagonal terms in Eq. (\ref{eq2-22}) are not the complex
conjugate of each other. So in general, there is no simple laser interaction leading
to Eq. (\ref{eq2-22}). That means constructing STA by nullifying all the non-adiabatic couplings
seems to be impracticable. Finding new supplementary Hamiltonian which is feasible in practice is imperative.
Under the premise that the off-diagonal terms in $H_{1}(t)$ are the complex
conjugate of each other, we assume the supplementary Hamiltonian is in the form of
\begin{eqnarray}\label{eq2-23}
  H_{1}(t)=\frac{\hbar}{2}\left(
                          \begin{array}{cc}
                            \delta_{+}(t) &\ \Omega(t) \cr\cr
                            \Omega^{*}(t) &\ \delta_{-}(t)
                          \end{array}
                    \right),
\end{eqnarray}
where $\delta_{\pm}$ and $\Omega$ are also undetermined coefficients.
Then, based on $H_{1}^{e}(t)=\tilde{R}^{\dag}H_{1}R$, we obtain
\begin{eqnarray}\label{eq2-24}
  H_{1}^{e}(t)=\frac{\hbar}{2}\left(
                       \begin{array}{cc}
                         \delta_{+}\cos^{2}{\frac{\theta}{2}}+\delta_{-}\sin^{2}{\frac{\theta}{2}}+\text{Re}[\Omega]\frac{\sin{\theta}}{2} \ \
                         & \frac{f_{-}}{f_{+}}(\delta_{+}-\delta_{-})\frac{\sin{\theta}}{2}-i\frac{f_{-}}{f_{+}}\text{Im}[\Omega]-\frac{f_{-}}{f_{+}}\text{Re}[\Omega]\cos\theta \cr\cr
                         \frac{f_{+}}{f_{-}}(\delta_{+}-\delta_{-})\frac{\sin{\theta}}{2}+i\frac{f_{+}}{f_{-}}\text{Im}[\Omega]-\frac{f_{+}}{f_{-}}\text{Re}[\Omega]\cos\theta  \ \
                         & \delta_{+}\sin^{2}{\frac{\theta}{2}}+\delta_{-}\cos^{2}{\frac{\theta}{2}}-\text{Re}[\Omega]\frac{\sin\theta}{2}
                       \end{array}
                    \right),
\end{eqnarray}
where $\text{Re}[*]$ and $\text{Im}[*]$ mean the real part and the
imaginary part of ``$*$'', respectively. Hence, under the condition
\begin{eqnarray}\label{eq2-25}
  (\delta_{+}-\delta_{-})\frac{\sin{\theta}}{2}-i\text{Im}[\Omega(t)]-\text{Re}[\Omega(t)]\cos\theta=i\partial_{t}{\theta},
\end{eqnarray}
the term in line 1 column 2 of $H_{0}^{e}$ will be nullified. Then,
as long as the system is initially in the adiabatic basis $|\phi_{+}(t)\rangle$ corresponding to $g_{+}(t_{0})=1$ and $g_{-}(t_{0})=0$,
it will remain in $|\phi_{+}(t)\rangle$ all the time.
The general solution of Eq. (\ref{eq2-25}) is
\begin{eqnarray}\label{eq2-25b}
  \text{Re}[\partial_{t}{\theta}]&=&\text{Im}[\lambda(t)]-\text{Im}[\Omega(t)]-\text{Im}[\zeta(t)], \cr\cr
  \text{Im}[\partial_{t}{\theta}]&=&-\text{Re}[\lambda(t)]+\text{Re}[\zeta(t)],
\end{eqnarray}
where $\lambda(t)=\frac{1}{2}(\delta_{+}-\delta_{-})\sin{\theta}$ and $\zeta(t)=\text{Re}[\Omega(t)]\cos\theta$.
Generally speaking, it does not matter whether the supplementary Hamiltonian is Hermitian or not.
If it is necessary to reduce the pulse intensity of the supplementary Hamiltonian, according to Eq. (\ref{eq2-25b}),
the simplest operation is increasing the imaginary part of $\lambda(t)$. In this case,
$(\delta_{+}-\delta_{-})$ should be a complex function in order to make $\lambda$'s imaginary part to be controllable.
In a limiting case that $\text{Im}[\lambda(t)]=\text{Re}[\partial_{t}{\theta}]$ and $\text{Re}[\lambda(t)]=-\text{Im}[\partial_{t}{\theta}]$,
$\Omega=0$, that is, we can speed up the adiabatic process without increasing the coupling intensity.
However, such case requires a very precise task on the imaginary part of $\delta_{\pm}$ (the imaginary part of $\delta_{\pm}$ could be
regarded as a supplementary decay rate or a dephasing rate), which increases the experimental complexity.

\begin{figure}
 \scalebox{0.35}{\includegraphics {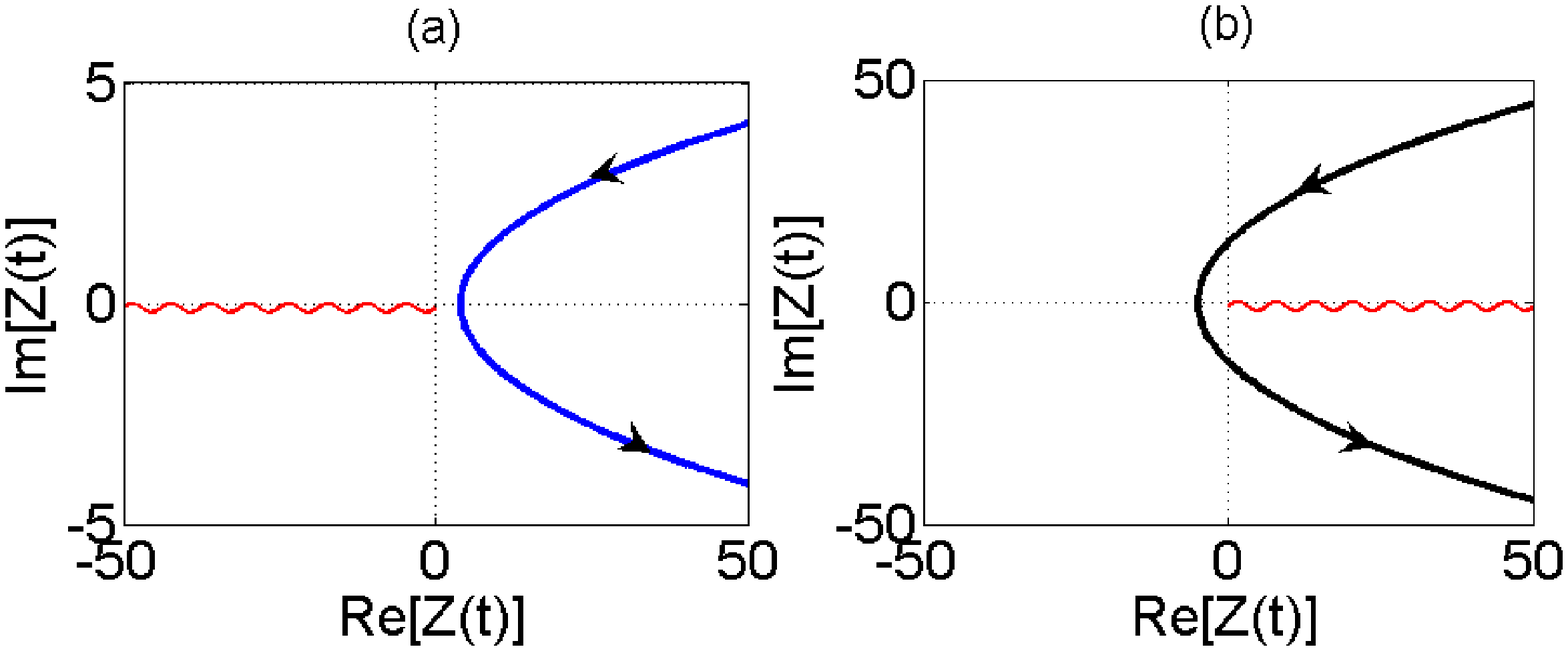}}
 \caption{
         Branch cuts and representative trajectories in the complex $Z(t)$ planes.
         For (a), $\gamma<2\Omega_{0}$, parameters are $\{t_{f}=\tau,\ \Omega_{0}=1/\tau,\ \Delta_{0}=9/\tau,\ \gamma=0.3/\tau\}$;
         For (b), $\gamma>2\Omega_{0}$, parameters are $\{t_{f}=\tau,\ \Omega_{0}=1/\tau,\ \Delta_{0}=9/\tau,\ \gamma=3/\tau\}$.
         In (a) the branch cut, just below the negative real axis, is chosen so that $-\pi\leq\eta\leq\pi$;
         in (b), just below the positive real axis, is chosen so that $0\leq\eta\leq 2\pi$.
         }
 \label{fig00}
\end{figure}

\begin{figure}
 \scalebox{0.19}{\includegraphics {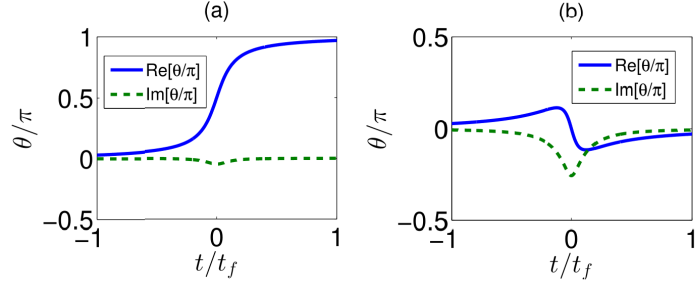}}
 \caption{
         The complex angle $\theta$ versus time.
         For (a), $\gamma<2\Omega_{0}$, parameters are $\{t_{f}=\tau,\ \Omega_{0}=1/\tau,\ \Delta_{0}=9/\tau,\ \gamma=0.3/\tau\}$;
         For (b), $\gamma>2\Omega_{0}$, parameters are $\{t_{f}=\tau,\ \Omega_{0}=1/\tau,\ \Delta_{0}=9/\tau,\ \gamma=3/\tau\}$.
         }
 \label{fig01}
\end{figure}

Without loss of generality,
we would like to take a Hermitian supplementary Hamiltonian (the easiest one to be realized in practice) as an example in the following discussion.
In this case, $\delta_{\pm}$ are real and the general solution of Eq. (\ref{eq2-25}) is (we assume Re$[\Omega(t)]=0$ for simplicity)
\begin{eqnarray}\label{eq2-26}
  \Omega_{a}=-\text{Re}[\partial_{t}{\theta}]+(\delta_{+}-\delta_{-})\frac{\text{Im}[\sin{\theta}]}{2}, \ \
  \delta_{+}-\delta_{-}=-\frac{2\text{Im}[\partial_{t}{\theta}]}{\text{Re}[\sin{\theta}]},
\end{eqnarray}
where $\Omega_{a}=\text{Im}[\Omega(t)]$.
Then, with $g_{+}(t_{0})=1$ and $g_{-}(t_{0})=0$, the solution of the Schr\"{o}dinger equation $i\hbar\partial_{t}|\psi^{e}(t)\rangle=H^{e}(t)|\psi^{e}(t)\rangle$, where
$H^{e}(t)=H_{0}^{e}(t)+H_{1}^{e}(t)$, is
\begin{eqnarray}\label{eq2-27}
  \left(
       \begin{array}{c}
         g_{+}(t) \cr\cr
         g_{-}(t)
       \end{array}
  \right)
  =
  \left(
       \begin{array}{c}
         \exp[{-i\int_{t_{0}}^{t}\frac{E_{+}(t')}{\hbar}-\frac{i\partial_{t}{f}_{+}(t')}{f_{+}(t')}}+\frac{\delta_{+}}{2}\cos^{2}{\frac{\theta(t')}{2}}+\frac{\delta_{-}}{2}\sin^{2}{\frac{\theta(t')}{2}}dt'] \cr\cr
         0
       \end{array}
  \right).
\end{eqnarray}
For simplicity, we can set $\delta_{+}=-\delta_{-}=\delta$, and Eq. (\ref{eq2-27}) could be
simplified as
\begin{eqnarray}\label{eq2-28}
  \left(
       \begin{array}{c}
         g_{+}(t) \cr\cr
         g_{-}(t)
       \end{array}
  \right)
  =
  \left(
       \begin{array}{c}
         \exp[{-i\int_{t_{0}}^{t}\frac{E_{+}(t')}{\hbar}-\frac{i\partial_{t}{f}_{+}(t')}{f_{+}(t')}}+\frac{\delta\cos{\theta}}{2} dt'] \cr\cr
         0
       \end{array}
  \right).
\end{eqnarray}
Obviously, the condition for $|g_{+}(t)|=1$ is
\begin{eqnarray}\label{eq2-29}
  \text{Im}[\frac{E_{+}(t')}{\hbar}-\frac{i\partial_{t}{f}_{+}(t')}{f_{+}(t')}+\frac{\delta\cos{\theta}}{2}]=0,
\end{eqnarray}
leading to
\begin{eqnarray}\label{eq2-29b}
  f_{+}(t)=\exp\{{\int_{t_{0}}^{t}{\text{Im}[\frac{E_{+}(t')}{\hbar}-\frac{\text{Im}[\partial_{t}{\theta}(t')]\cdot\cos{\theta(t')}}{2\text{Re}[\sin{\theta(t')}]}]}dt'}\}.
\end{eqnarray}
Returning Eq. (\ref{eq2-28}) to the interaction frame with relationship $|\psi(t)\rangle=R(t)|\psi^{e}(t)\rangle$, the evolution state is
\begin{eqnarray}\label{eq2-30}
  |\psi(t)\rangle&=&
            \left(
                 \begin{array}{cc}
                   f_{+}(t)\cos{\frac{\theta}{2}} &\ f_{-}(t)\sin{\frac{\theta}{2}} \cr\cr
                   f_{+}(t)\sin{\frac{\theta}{2}} &\ -f_{-}(t)\cos{\frac{\theta}{2}}
                 \end{array}
            \right)\cdot
            \left(
               \begin{array}{c}
                 g_{+}(t) \cr\cr
                 g_{-}(t)
            \end{array}
  \right) \cr\cr\cr
  &=&f_{+}(t)g_{+}(t)\left(
       \begin{array}{c}
         \cos\frac{\theta}{2} \cr\cr
         \sin\frac{\theta}{2}
       \end{array}
  \right).
\end{eqnarray}

\begin{figure}
 \scalebox{0.4}{\includegraphics {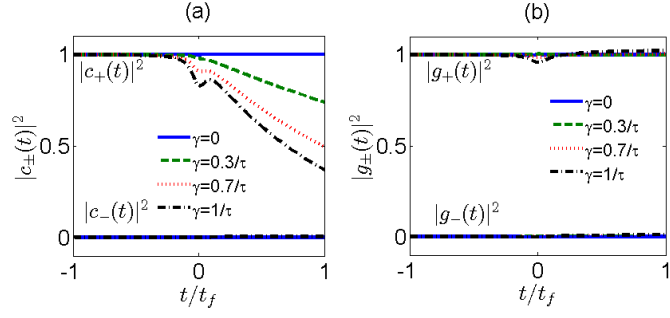}}
 \caption{
         (a) $|c_{+}(t)|^{2}$ versus time; (b) $|g_{+}(t)|^{2}$ versus time.
         In an Allen-Eberly process with different decay rates $\gamma$ when $\{t_{f}=\tau,\ \Omega_{0}=1/\tau,\ \Delta_{0}=9/\tau\}$ .
         }
 \label{fig3}
\end{figure}

\begin{figure}
 \scalebox{0.35}{\includegraphics {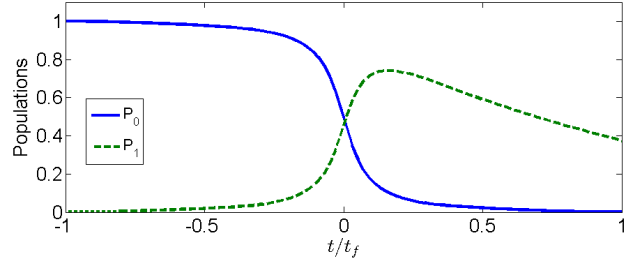}}
 \caption{
         Time-dependent populations for bare states $|0\rangle$ and $|1\rangle$ defined by $P_{m}=|\langle m|\psi(t)\rangle|^{2}$.
         Parameters are $\{t_{f}=\tau,\ \Omega_{0}=1/\tau,\ \Delta_{0}=9/\tau,\ \gamma=1/\tau\}$.
         }
 \label{fig4}
\end{figure}

We study now the laser-driven coherent decay from the
excited state of a two-level system with slow spontaneous
decay. This type of decay process is of interest as it occurs
coherently, unlike the incoherent spontaneous emission.
We will apply the present method to the famous Allen-Eberly model to verify the validity of the optimized method.
For an Allen-Eberly model, the Rabi frequency $\Omega_{R}(t)$ and detuning $\Delta(t)$ are given as
\begin{eqnarray}\label{eq3-13}
  \Omega_{R}(t)=\Omega_{0}\text{sech}({t}/{\tau}),\ \Delta(t)=\Delta_{0}\text{tanh}(t/\tau),
\end{eqnarray}
where $\Omega_{0}$ is the pulse amplitude, $\Delta_{0}$ corresponds to the chirp
rate, and $\tau$ is the characteristic duration of the interaction.
Once $\Omega_{R}(t)$ and $\Delta(t)$ are specified, we should firstly analyze the behavior of the radicand in
Eq. (\ref{eq2-2}),
\begin{eqnarray}\label{eq3-4a}
  Z(t)=-[\gamma+2i\Delta(t)]^{2}+4\Omega_{R}^{2}(t),
\end{eqnarray}
where $Z(t)$ is in polar form $Z=\lambda e^{i\eta}$, with modulus $\lambda=|\sqrt{\text{Re}[Z(t)]^{2}+\text{Im}[Z(t)]^{2}}|$ and argument $\eta$.
According to Eq. (\ref{eq2-4}), two regimes can be distinguished for this protocol depending on $\gamma>2\Omega_{0}$
and $\gamma<2\Omega_{0}$ (a degeneracy exists at $t=0$ if $\gamma=2\Omega_{0}$).
(1) When $\gamma<2\Omega_{0}$, then $\text{Re}[Z(t)]>0$. A
representative trajectory of $Z(t)$ in the complex $Z(t)$ plane is
shown in Fig. \ref{fig00} (a) with parameters $\{t_{f}=\tau,\ \Omega_{0}=1/\tau,\ \Delta_{0}=9/\tau,\ \gamma=0.3/\tau\}$.
The branch cut of the square
root just below the negative real axis is chosen so that
$-\pi<\eta<\pi$.
We depict the trajectory of the complex angle $\theta(t)$
in Fig. \ref{fig01} (a), which shows $\theta(t_{0})\approx
0$ and $\theta(t_{f})\approx \pi$. In this case, the eigenvector
$|+(t)\rangle$ will evolve from $|+(t_{0})\rangle=|0\rangle$ to
$|+(t_{f})\rangle=|1\rangle$.
(2) When $\gamma>2\Omega_{0}$,
$Z(t)$ crosses the negative real axis as shown in Fig. \ref{fig00} (b).
We also depict the form of the $\theta$ trajectory in Fig. \ref{fig01} (b).
As we can find from the figure, $\theta$ changes very slightly during the whole process,
while such a change in $\theta$ is not enough to realize a desired population transfer.
Parameters are chosen as $\{t_{f}=\tau,\ \Omega_{0}=1/\tau,\ \Delta_{0}=9/\tau,\ \gamma=3/\tau\}$ in plotting Figs. \ref{fig00} (b) and (d).

In the following, we will discuss the validity of the present non-Hermitian STA  by
examples of fast population inversion.
$|c_{\pm}(t)|^{2}$ and $|g_{\pm}(t)|^{2}$ which can be regarded as the traditional population and modified population for the eigenvectors $|\pm(t)\rangle$, respectively,
are displayed in Fig. \ref{fig3} with different decay rates ($\gamma<2\Omega_{0}$).
Figure \ref{fig3} (a) shows that $|c_{+}(t)|^{2}$ decays strongly, especially, when
$\gamma$ is relatively large. As a contrast,
$|g_{+}(t)|^{2}$ is approximate to $1$ all the time, and
it changes slightly along with the increasing of $\gamma$.
Meanwhile, $|c_{-}(t)|^{2}$ (or $|g_{-}(t)|^{2}$) remains zero all the time
which demonstrates the eigenvector $|-(t)\rangle$ is unpopulated.
Nevertheless, it is still a little different from our expectation that $|g_{+}(t)|^{2}$ does not ideally equal to $1$ when $\gamma$ is relatively large.
The difference is due to the fact that the system is not exactly in $|+(t)\rangle$ at initial time with a relatively large $\gamma$.
Then, we plot time-dependent populations [Fig.
\ref{fig4}] for bare states $|0\rangle$ and $|1\rangle$, with
$\gamma=1/\tau$ (other parameters are the same as that in plotting Fig. \ref{fig00}). As shown in the figure, the
fast population inversion for a non-Hermitian system is realized (for $t\rightarrow t_{f}$, we have $P_{0}=0$ and $P_{1}\neq 0$).

\section{Conclusion}
Non-Hermitian system has its natural advantages than Hermitian
system in speeding up a slow adiabatic passage \cite{Pra87042332,Qip13371}.
The defects of the previous non-Hermitian STAs, however,
restrict the applications of speed-up schemes in non-Hermitian
systems to a certain extent. Therefore, in light of Refs.
\cite{Pra84023415} and \cite{Pra8705250289063412}, we have proposed
an effective method to improve non-Hermitian STA for a better application in quantum information processing.
This method is performed by introducing a series of redesigned supplementary
Hamiltonians to nullify the specified non-adiabatic couplings so that the evolution of the
system would be confined in the reference instantaneous eigenstate.
In this way, the present method allows one to speed up
a non-Hermitian adiabatic process with arbitrary decay rate by using a Hermitian supplementary Hamiltonian.
Hence, realizing non-Hermitian STA could be much easier in practice.
Moreover, we have applied this method to the Allen-Eberly model
and shown with numerical simulation that the ultrafast population inversion could be determinately
achieved in a two-level non-Hermitian system.



Assessing the cost of implementing STA arises as a
natural question with both fundamental and practical
implications in nonequilibrium statistical mechanics.
For unitary systems, thermodynamic cost by using STA has been discussed in Refs. \cite{Prl118100601,Prl118100602,Prl110050403,Pra96022133}, while
for non-unitary systems, the time-energy cost and quantum speed-limit are still worth to be studied.
Therefore, exploring the time-energy cost and quantum speed-limit for the non-Hermitian STA is an interesting subject in the future work.

\section*{ACKNOWLEDGMENTS}
It is a pleasure to thank Dr. Z.C. Shi for valuable discussions.
This work was supported by the National Natural Science Foundation
of China under Grants No. 11575045, No. 11374054 and No. 11675046,
and the Major State Basic Research Development Program of China
under Grant No. 2012CB921601.

\end{document}